\documentstyle[12pt]{article}
\begin{document}
\normalbaselineskip=24 true pt
\normalbaselines
\bibliographystyle{unsrt}

\def\be {\begin{equation}}
\def\ee {\end{equation}}
\def\bea {\begin{eqnarray}}
\def\eea {\end{eqnarray}}
\def\b {\bibitem}
\def\tw {\theta_W}
\def\r {\rightarrow}
\def\zbb {Z\rightarrow b\bar b}
\def\fcfr {f_k^cv_k^c+f_k^rv_k^r}
\def\gcgr {g_k^cv_k^c+g_k^rv_k^r}
\thispagestyle{empty}

\begin{flushright}
{\large\sf SINP-TNP/95-10}\\
{\large\sf MRI-PHY/15/95}\\
{\large\sf July 1995}
\end{flushright}
\vskip 0.5 true cm
\begin{center}
{\Large\bf A General Higgs Sector:\\
Constraints and Phenomenology}
\end{center}
\vskip 0.5 true cm
\begin{center}
{\large\sf Anirban Kundu}\\
Theory Group, Saha Institute of Nuclear Physics,\\
1/AF Bidhannagar, Calcutta - 700 064, India.\\
E-mail: akundu@saha.ernet.in\\[5mm]
and\\[5mm]
{\large\sf Biswarup Mukhopadhyaya}\\
Mehta Research Institute,\\
10 Kasturba Gandhi Marg, Allahabad - 211 002, India.\\
E-mail: biswarup@mri.ernet.in
\end{center}
\newpage
\begin{abstract}

We have investigated some phenomenological aspects of an $SU(2)\times U(1)$
scenario where scalars belonging to arbitrary representations of $SU(2)$
are involved in electroweak symmetry breaking. The resulting interaction
terms are derived. Some constraints are obtained on the arbitrary scalar
sector from the requirement of tree-level unitarity in longitudinal
gauge boson scattering. We also show that, in cases where the scalars ensure
$\rho=1$ at tree-level, useful restrictions on their parameter space
follow from precision measurements of the $Zb\bar b$ vertex. Finally, some
salient features about the production of such Higgs bosons
in $e^+e^-$ collision are discussed.

\end{abstract}
\newpage

\centerline{\bf 1. Introduction}
\bigskip

The electroweak symmetry breaking sector of the Standard Model (SM)
is still an object of widespread conjectures. Admittedly, the model
with one Higgs doublet is the simplest one and is also consistent
with the current experimental results. From an alternative standpoint,
however, it may seem to be rather artificial to postulate just {\em one}
fundamental scalar doublet in nature, as compared with several families
of particles in the fermionic sector. Models with two or more doublets
have been explored in this spirit \cite{hunter}.

It is also pertinent to investigate the consequences of scalars belonging
to other representations of $SU(2)$. This will not alter the gauge group
structure of the SM, but will enlarge its particle content and change
the gauge-scalar and fermion-scalar interactions in a significant manner.
The motivation for introducing higher Higgs representations can be seen,
for example, in the context of neutrino masses. If lepton number can be
violated, then the interaction with the vacuum expectation value (VEV)
of a Higgs triplet may lead to Majorana masses for
left-handed neutrinos \cite{gelmini}.
This not only facilitates the incorporation of neutrino masses without
any need for completely sterile right-handed neutrinos, but also explains why
neutrino masses are so much smaller than their charged fermionic
counterparts.
Triplet Higgs scenarios have been further explored in the literature
in both the contexts of $e^+e^-$ and hadronic colliders \cite{gunion,
vega,hontka,roger,lusignoli,swartz,bamert}.

However, representations of scalars higher than doublets suffer from one
serious malady, namely, they in general violate the experimental constraints
on the $\rho$-parameter, defined as $\rho=m_W^2/ m_Z^2\cos^2\tw$.
The tree-level value of $\rho$ in a general Higgs scenario is given by
\cite{hunter}
\be
\rho={\sum_k[4T_k(T_k+1)-Y_k^2]|v_k|^2c_k\over \sum_k2Y_k^2|v_k|^2}
\ee
where $v_k$ is the VEV of the $k$-th multiplet of scalars, and $c_k=1$(1/2)
for a complex (real) representation.
It can be verified from above that higher scalar representations contradict
the experimentally measured value of $\rho=1.0004\pm 0.0030$ \cite{pdg}
unless one of
the following conditions is satisfied:

\begin{enumerate}

\item We have some higher Higgs representation which {\em accidentally}
does not contribute to $\rho$, such as one with $T=3$, $Y=4$.

\item The VEVs of the higher representations are much smaller than the
doublet VEV so that $\rho$ is not affected significantly.

\item There is some custodial symmetry among the higher representations
such that their contributions to $\rho$ cancel each other.

\end{enumerate}

It should be noted that although the last possibility smacks of fine-tuning,
especially where higher order effects are taken into account, some serious
model-building has been done in recent times on its basis \cite{georgi1,
georgi2,golden}.

Based upon the experiences summarised above, we attempt in this paper
to discuss the state of affairs in a general scenario with an arbitrary
collection of scalar representations, both real and complex. We have also
tried to see if constraints other than those from the $\rho$-parameter
can be imposed on such a scenario. It will be shown below, for example,
that the experimental measurement of the effective $Zb\bar b$ vertex
can lead to restrictions on an arbitrary Higgs structure (if it is not
already constrained by the $\rho$-parameter), which are comparable with,
and sometimes more stringent than, those on two-Higgs doublet models.
That the longitudinal gauge boson scattering amplitudes should respect
unitarity at high energies also lead to some nontrivial constraints
on the scalar sector.

In Section 2 we discuss the various interactions in the Lagrangian with
a general assortment of Higgs multiplets. The constraints on them arising
from the requirement of tree-level unitarity in longitudinal gauge boson
scattering are derived in Section 3. In Section 4 we take up the limits from
the $Zb\bar b$ effective vertex. Some further phenomenological implications,
like Higgs production by the Bjorken process and the effects of an $H^{\pm}
W^{\mp}Z$  tree-level interaction, are discussed in Section 5. We conclude
our discussions in Section 6, and follow up with some of the formulae
listed in the Appendices.

\bigskip\bigskip

\centerline{\bf 2. Formalism}

\bigskip

A scalar multiplet in the $SU(2)\times U(1)$ context can occur in both
real ($Y=0$) and complex ($Y\not = 0$) representations. In general, the
Lagrangian in the scalar sector contains both the kinetic energy and
the potential terms. The potential, of course, has to depend on the
particle content of specific models. Since our purpose here is to investigate
the general or model-independent features of an arbitrary Higgs structre,
we wish to refrain from attaching ourselves to any particular form of the
potential. The general potential consists of all possible gauge-invariant
quadratic and quartic terms formed out of all the fields present in the
scenario. The imposition of some additional symmetry or phenomenological
requirement eliminates or relates some of these terms. It should be noted that
any particular choice of the potential determines the scalar self-couplings
and the mixing matrices relating the weak eigenstates to the physical
(or mass) eigenstates.

The kinetic term of the Higgs fields, which also includes the
interaction terms with the electroweak gauge fields by virtue of the
covariant derivative, can be written as
\be
{\cal L}_{kin}=\sum_k[(D_{\mu}\Phi_k^c)^{\dag}(D^{\mu}\Phi_k^c)
+{1\over 2}(D_{\mu}\Phi_k^r)^T(D^{\mu}\Phi_k^r)],
\ee
where $\Phi^c(\Phi^r)$ denotes a complex (real) Higgs multiplet,
and a sum over repeated Greek indices is implied. The explicit
form of the covariant derivative is
\be
D_{\mu}=\partial_{\mu}+ig \sum_{i=1}^3W_{\mu}^iT_i + {i\over 2}g'YB_{\mu},
\ee
where
\be
T^{\pm}=T_1\pm iT_2
\ee
and $T_3$ are the $SU(2)$ generators of proper dimension.

Here let us clarify our notations. We will denote the Higgs fields,
both complex and real, in the
weak basis by $\phi$ (and the multiplets by $\Phi$)
and those in the mass basis by $H$. These two
sets will be related by unitary matrices. The neutral scalar and
pseudoscalar fields do not mix with each other; thus, we have two distinct
mixing matrices in the neutral sector. All Higgs fields with same nonzero
charge can mix with each other. We assume these mixing elements to be real
so that there is no $CP$ violation in the Higgs sector. This does not affect
the generality of our treatment since we will focus on $CP$ conserving
phenomena. Thus, we can have the following relations between the fields
in the weak and the mass basis:
\bea
H_i^{\pm} &=& \alpha_{ij}\phi_j^{\pm},\ \ \phi_i^{\pm}=\alpha_{ji}H_j^{\pm},
\nonumber\\
H_i^{\pm\pm} &=& \sigma_{ij}\phi_j^{\pm\pm},\ \ \phi_i^{\pm\pm}=\sigma_{ji}
H_j^{\pm\pm}, \nonumber\\
H_{Si}^{0} &=& \beta _{ij}\phi_{Sj}^{0},\ \ \phi_{Si}^{0}=\beta _{ji}H_{Sj}^0,
\nonumber\\
H_{Pi}^{0} &=& \gamma_{ij}\phi_{Pj}^{0},\ \ \phi_{Pi}^{0}=\gamma_{ji}H_{Pj}^0,
\eea
and similarly other relations between the triple- and higher-charged Higgs
fields can be written. As we have mentioned before, the specific forms of the
mixing matrices $\alpha, \beta$ etc. depend on the potential. The elements
of the first row of the
$\alpha$ and $\gamma$ matrices are fixed from the definition of
the Goldstone bosons, which of course include the VEVs of the scalar fields
which are in its turn, functions of the potential. We denote the charged
Goldstone bosons $G^{\pm}$ as $H_1^{\pm}$, and the neutral Goldstone
boson $G^0$ as $H_{P1}^0$. The subscripts $S$ and $P$ denote the scalar
and the pseudoscalar states respectively.

The Goldstone states can directly be found out from the Lagrangian
by looking at the non-diagonal pieces (which are cancelled by the
gauge-fixing term). The expressions for the Goldstones immediately
follow:
\bea
G^+&=&{g\over\sqrt{2}m_W}\sum_k[(T^+v_k)^{\dag}\phi_k-\phi_k
^{\dag}(T^-v_k)],\\
G^-&=&{g\over\sqrt{2}m_W}\sum_k[\phi_k^{\dag}(T^+v_k)-
(T^-v_k)^{\dag}\phi_k],\\
G^0&=&{ig\over 2\cos\theta_Wm_Z}\sum_k[\phi_k^{\dag}Y_kv_k-
v_kY_k\phi_k],
\eea
which are normalized to the SM expressions. The sum here runs over both
the real as well as the complex representations. However, $G^0$ gets
contribution only from the complex representations, as $Y_{real}=0$.
Also, one has to introduce a factor
of $1/2$ for real representations as both
the particle and the antiparticle states are in the same multiplet.

If we make the additional assumption that for all complex representations,
$T^-v=0$, i.e., the neutral member of the multiplet has the lowest
weight, then we have $T=Y/2$ for all such representations. Such a choice
includes, without any loss of generality, all doublets as well as the
triplet models which are of current phenomenological interest
\cite{hunter}. It simplifies
the form of the charged Goldstone bosons given above in the sense that
$T^-v_k$ terms will be absent:
\bea
G^+&=&{g\over \sqrt{2}m_W}\sum_{k=1}^n[\sqrt{n_k-1}v^c_k+{1\over 2}
\sqrt{n_k^2-1}v^r_k]\phi_k^+\nonumber\\
&=&\sum_k\alpha_{1k}\phi_k^+.
\eea
Here, $v^c_k$ ($v^r_k$) is the VEV of the $k$-th complex (real)
representation, whose dimension is $n_k$. Note that one row of the
matrix $\alpha$ gets determined in this way.

With $T^-v=0$, the gauge boson masses are given by
\bea
m_{W^{\pm}}^2&=&{1\over 2}g^2\sum_k[T(T+1)(v^r_k)^2+2T(v_k^c)^2]\\
m_Z^2&=&{1\over 2}g^2\sum_k 4T^2\sec^2\theta_W(v_c^k)^2.
\eea
Thus, the real multiplets do not contribute to $m_Z$, whereas both the
real as well as the complex multiplets do contribute to $m_W$. This also
recasts eq. (1) into
\be
\rho={\sum_k\big[
T(T+1)(v^r_k)^2+2T(v^c_k)^2\big]\over\sum_k 4T^2(v_k^c)^2}.
\ee
Thus, a good way to have the custodial $SU(2)$ symmetry intact with more
than one `bad' representations is the following. We can have one or
more doublets or singlets with arbitrary VEVs; they do not affect
$\rho$. Next, we add one complex $n$-plet ($T=Y/2$) with neutral VEV
$=b$, and ${4T-2\over T+1}$ (if integer) number of real $n$-plets
($T=n, \ Y=0$) with same VEV. However, if ${4T-2\over T+1}$ is not an
integer, one has to add a single real $n$-plet (for example) with VEV
$=({4T-2\over T+1})^{1/2}b$. (Note that this option is available also
when ${4T-2\over T+1}$ is an integer.)
For $T=1$, this prescription reproduces
the triplet Higgs model of Georgi and Machacek \cite{georgi2}.
For $T=2$, we have a
complex and two real 5-plets with same VEV, which respect a custodial
$SU(2)$ symmetry. This will henceforth be called the 5-plet model.

The forms for the vertex factors directly follow from eq. (2). We show
some typical vertices which will be directly relevant in our future
discussions. Appendix 1 contains a more
complete list. Here, we adopt the simplification
that for a complex representation $T^-v_k=0$.

First, let us show some couplings involving two gauge bosons and one scalar.
The $ZZ\phi^0_{Sk}$ vertex factor is given by ${ig^2\over\sqrt{2}
\cos^2\tw}v_kY_k^2g_{\mu\nu}$. (The $ZZ\phi^0_{Pk}$ term is forbidden
from parity conservation.) In the mass basis, this factor can be written
as
\be
ZZH^0_{Si}\ :\ {ig^2g_{\mu\nu}\over \sqrt{2}\cos^2\tw}\sum_k\beta_{ik}
v_kY_k^2
\ee
which is obvious from eq. (5). The remaining interactions are given below in
the mass basis only. For example,
\bea
W^+W^-H^0_{Si}\ &:&\ {ig^2g_{\mu\nu}\over\sqrt{2}}\sum_k\beta_{ik}
(v^c_kY_k+{1\over 4}(n_k^2-1)v^r_k)\\
W^+ZH^-_{i}\ &:&\ {ig^2g_{\mu\nu}\over\sqrt{2}\cos\tw}\sum_k\alpha_{ik}
(f^c_kv^c_k+f^r_kv^r_k)
\eea
where
\be
f^c_k=\sqrt{n_k-1}(\cos^2\tw-Y_k);\ \ f^r_k={1\over 2}\sqrt{n_k^2-1}
\cos^2\tw.
\ee
Another interesting coupling, which occurs with at least real 5-plets
or complex triplets, is
\be
W^+W^+H^{--}_{i}\ :\ {ig^2g_{\mu\nu}\over{2}}\sum_k\sigma _{ik}
(g^c_kv^c_k+g^r_kv^r_k)
\ee
with
\bea
g^c_k&=&\sqrt{2(n_k-1)(n_k-2)}\\
g^r_k&=&{1\over 4}\sqrt{(n_k^2-1)(n_k^2-9)}.
\eea
As emphasized earlier, the expression for $g_c^k$ does depend on the
approximation $T^-v_k=0$; however, $g^r_k$ is independent of any such
approximation and uses only the fact that $Y_{real}=0$.

In the next category for two scalar-one gauge boson couplings, we only
show the $ZH_i^+H_j^-$ vertex:
\be
ZH_i^+H_j^-\ : \ -{ig\over 2\cos\tw}(p_1+p_2)_{\mu}[2\cos^2\tw\delta_{ij}
-\sum_k\alpha_{ik}\alpha_{jk}Y_k]
\ee
where $p_1$ is incoming and $p_2$ is outgoing at the vertex. From this,
the expressions for vertices involving one or two charged Goldstones
follow immediately:
\bea
ZH_i^+G^-\ &:& \ -{ig\over 2\cos\tw}(p_1+p_2)_{\mu}[2\cos^2\tw\delta_{i1}
-\sum_k\alpha_{ik}\alpha_{1k}Y_k],\\
ZG^+G^-\ &:& \ -{ig\over 2\cos\tw}(p_1+p_2)_{\mu}\sum_k\alpha_{1k}^2
(2\cos^2\tw -Y_k).
\eea
It may be noted that at the proper limit, the couplings corresponding
to the SM or the two-Higgs doublet model are successfully reproduced.
For a general representation, elements of the mixing
matrices depend on the chosen form of the potential.
However, we will show that one can extract some model-independent
information about them.

{}From the expressions given above (and also in Appendix 1) one observes
that {\em all scalars} (except a singlet) in general couple
to the electroweak gauge bosons. On the other hand, only weak doublets
couple with fermions; that too in a restricted manner to avoid
unreasonably large flavour-changing neutral currents (FCNC).
Following the conditions of natural flavour conservation \cite{glashow},
we will invoke two kinds of
models: in the first one, only one doublet (let us
call it $\Phi_1$) couples to both up- and down-type quarks (model 1) and
in the second one, $\Phi_1$ couples with the up-type and $\Phi_2$ with the
down-type quarks (model 2). Other doublets, if present, do not couple
to the fermions \cite{grossman}.

In model 1, the Yukawa couplings are as follows:
\bea
\bar u d H_i^+\  &:&\  {ig\over \sqrt{2}m_W}{\alpha_{i1}\over\alpha_{11}}
(m_uP_L-m_dP_R),\\
\bar u d G^+\  &:&\  {ig\over \sqrt{2}m_W}
(m_uP_L-m_dP_R).
\eea
In model 2, their counterparts are
\bea
\bar u d H_i^+\  &:&\  {ig\over \sqrt{2}m_W}({\alpha_{i1}\over\alpha_{11}}
m_uP_L-{\alpha_{i2}\over\alpha_{12}}m_dP_R),\\
\bar u d G^+\  &:&\  {ig\over \sqrt{2}m_W}
(m_uP_L-m_dP_R),
\eea
where
\be
P_L={1-\gamma_5\over 2},\ \ P_R={1+\gamma_5\over 2}.
\ee
Note that no sum over weak eigenstates appears as only one weak doublet
is responsible for giving mass to a particular type of quark. The $\alpha$'s
in the denominator follow from the normalisation of charged Goldstones.
It also follows that for the $t-b$ system, where $m_t\gg m_b$, both models
yield the same $\bar t b H_i^+$ vertex.

Before concluding this section, we give some specific examples of the scalar
mixing matrices. An illustrative case is the triplet
model, consisting of a complex
($Y=2$) and a real triplet in addition to the standard doublet, where the
tree-level VEV's of the complex and the real triplet are taken to be equal to
maintain $\rho=1$ \footnote{It has already been discussed in the literature
\cite{gunion}
that maintaining this equality at higher orders requires fine-tuning at
a level comparable to that in the minimal SM.}.

In terms of the doublet-triplet mixing angle $\theta_H$, where $\tan\theta_H
=2\sqrt{2}b/a$ ($a/\sqrt{2}$ and $b$ being the doublet and the triplet
VEVs respectively), the mixing matrices $\alpha$, $\beta$ and $\gamma$
have the following form:
\be
\alpha=\pmatrix{c_H& s_H/\sqrt{2}& s_H/\sqrt{2}\cr o&1/\sqrt{2}& -1/
\sqrt{2}\cr -s_H& c_H/\sqrt{2}& c_H/\sqrt{2}},
\ee
\be
\beta=\pmatrix{1&0&0\cr 0&\sqrt{2/3}& 1/\sqrt{3}\cr 0&-1/\sqrt{3}&\sqrt{2/3}
},
\ee
\be
\gamma=\pmatrix{c_H& s_H\cr -s_H& c_H},
\ee
where
\bea
\phi^+_{1,2,3}&=& \phi^+, \chi^+, \xi^+,\\
H^+_{1,2,3}&=& G^+, H_5^+, H_3^+,\\
(\phi^0_S)_{1,2,3}&=&\phi^0_R,\chi^0_R,\xi^0,\\
(H_S^0)_{1,2,3}&=& H^0_1,{H'}^0_1,H_5^0,\\
(\phi^0_P)_{1,2}&=& \phi^0_I,\chi^0_I,\\
(H^0_P)_{1,2}&=& G^0,H_3^0\eea
the notations being the same as in Ref. \cite{gunion},
and no mixing between $H^0_1$
and ${H'}^0_1$ is assumed. This is necessary to fix the first row of the
$\beta$ matrix, as it cannot be fixed from any Goldstone normalisations.
{}From the forms of $\alpha$ and $\beta$, it is obvious that $H^{\pm}_5$,
$H^0_5$
and ${H'}^0_1$ do not couple to fermions. Also, it is easy to check, for
example,
that the structure of $\alpha$ rules out a $H_3^+W^-Z$ coupling.

Another example is the 5-plet model. It consists of one complex doublet of VEV
$a/\sqrt{2}$ and one complex 5-plet and two real 5-plets with VEV$=b$. As we
have already shown, such a choice automatically ensures $\rho=1$ at tree-level.
Defining $\tan\theta_H=2\sqrt{2}b/a$, we get, for example,
\be
\alpha=\pmatrix{c_H&s_H/2&\sqrt{3/8}s_H& \sqrt{3/8}s_H\cr
-s_H&c_H/2&\sqrt{3/8}c_H&\sqrt{3/8}c_H\cr
0&-\sqrt{3/4}&1/\sqrt{8}&1/\sqrt{8}\cr
0&0&1/\sqrt{2}&-1\sqrt{2}},
\ee
where
\bea
\phi^+_{1,2,3,4}&=&\phi^+(doublet),\ \ \chi_1^+(complex~5-plet),\nonumber\\
&{ }&\psi_1^+(real~5-plet),\ \ \psi_2^+(real~5-plet).
\eea
It is again obvious that $H_3^+$ and $H^+_4$ do not couple to
fermion-antifermion pairs.
\bigskip\bigskip
\newpage

\centerline{\bf 3. Unitarity Sum Rules}
\bigskip

The requirement of unitarity of the partial wave amplitudes for
longitudinal gauge boson scattering \cite{dawson}
can lead to some useful restrictions
on an arbitrarily extended scalar sector. On one hand, consideration
of the zeroth partial wave amplitudes at a centre-of-mass energy much higher
than the Higgs masses yields the upper limit of about 1 TeV \cite{lee} on
{\em at least} one scalar which interacts with a pair of gauge bosons.
On the other hand, one may demand that for arbitrarily large values of the
scalar masses, unitarity should hold for $\sqrt{s}~{<\atop\sim}~1$ TeV (i.e.,
ask for the same high-energy cut-off as that in the minimal SM). One
way of guaranteeing this is to impose the restriction that for each
scattering process, the total amplitude for all the Higgs-mediated
diagrams be equal to the minimal SM amplitude. We maintain that such
conditions are {\em sufficient} rather than necessary. However, they allow
us to relate and simplify the plethora of parameters in a scenario
containing an assortment of scalars. Here we present those among such
conditions which can be written in a model-independent way, i.e., without
recourse to the detailed form of the scalar potential. (For example,
computation of scattering processes involving the scalars either in the
initial or in the final channel requires scalar self-couplings, and hence
are omitted from our set of conditions.) As far as longitudinal gauge boson
scattering is concerned, all the relations that follow from the above criterion
can be obtained from two processes. We choose $W_LZ_L\r W_LZ_L$
and $W_L^+W_L^+\r W_L^+W_L^+$ for that purpose. All the other processes
can be easily seen to give the same set of conditions using crossing
symmetry.

The condition for the amplitude of $W_LZ_L\r W_LZ_L$ channel to
satisfy the unitarity bound is
\be
g^2\Big[\sum_kv_k^2(Y_k^3-f_k^2)+(\sum_k\alpha_{1k}f_kv_k)^2\Big]={1
\over 2}m_W^2
\ee
where the sum is over all multiplets, real and complex, and $f_k$ is the
proper factor ($f_k^c$ or $f_k^r$) as defined in eq. (16). The second term
in the left-hand side leaves out the Goldstone
contribution. Furthermore, in the left-hand side, the
term proportional to $Y_k^3$ comes from the $t$-channel graphs whereas the
one proportional to $f_k^2$ comes from the combined $s$ and $u$-channel
graphs; the latter two add with a negative sign to the $t$-channel diagram
to give the total contribution (as at $s,t,u\gg m_H^2, \ s+t+u\approx
0$) which should be equal to the SM contribution. $W^+_LW^-_L\r
Z_LZ_L$ gives the same condition from crossing symmetry.

The second condition comes from $W^+_LW^+_L\r W^+_LW^+_L$, which is related to
$W^+_LW^-_L \r W^+_LW^-_L$ by crossing. Here, doubly charged scalars in
an extended scenario play a significant role; however, Goldstone contributions
need not be subtracted as neutral pseudoscalars do not couple to $W$-pairs
from $CP$-invariance of the Lagrangian. The condition is
\be
{g^2\over 2}\sum_k\Big[\big(Y_k^2-{1\over 2}(g^c_k)^2\big)
(v^c_k)^2+\{ {1\over 4}
(n_k^2-1)-{1\over 2}(g^r_k)^2\}(v^r_k)^2\Big]=m_W^2.
\ee
Here $g^c_k$ and $g^r_k$ are those given in eqs. (18) and (19). The dependence
on the $\beta$- and $\sigma$-matrices cancel from completeness.

In processes like $f\bar f\r V_LV_L$ ($V=W,Z$), good high-energy
behaviour of the cross-section is ensured by gauge invariance. The same
situation holds with a complicated Higgs structure if one demands that
the sum of the amplitudes for all scalar-mediated diagrams be equal to the
SM Higgs-induced amplitude. However, it is straightforward to see that
such an equality is automatically ensured from the unitarity of the
$\beta$-matrix and the expression for $\alpha_{1i}$ in terms of
the scalar VEVs.
On the other hand, a similar requirement for the process $f_1\bar f_2
\r W_LZ_L$ yields a nontrivial restriction. It involves the
$H^{\pm}W^{\mp}Z$ couplings present in a general scenario. A simple
calculation gives the following sufficient condition:
\be
{f_1^cv_1\over \alpha_{11}}=\sum_k\alpha_{1k}(f^c_kv^c_k+f^r_kv^r_k).
\ee
If, in addition, the doubly charged scalars have $\Delta L=2$ couplings with
a pair of leptons, then a constraint on the doubly charged sector can be
obtained from processes like $e^-e^-\r W_L^-W_L^-$ \cite{swartz}.
\bigskip\bigskip

\centerline{\bf 4. Constraints from $Z\r b\bar b$}
\bigskip

Since the advent of microvertex detectors, the decay $\zbb$ \cite{hollik}
can be
tracked down with increasingly higher precision, and the latest results from
LEP-1 seriously encourage the view that there may be new physics
beyond the SM.
The reason is that the experimentally measured value
of the ratio $R_b$, defined as
\be
R_b={\Gamma (\zbb )\over \Gamma (Z\r {\rm hadrons}},
\ee
comes out to be nearly 2.2$\sigma$ above the SM value; the experiments
give $0.2202\pm 0.0020$ while the SM prediction is $0.2156\pm 0.0004$
\cite{erler},
for $m_t=175$ GeV. With the top quark mass more or less accurately known,
$R_b$ can act as a `new physics meter', as most of the QCD corrections
to the individual decay widths cancel in the ratio \cite{pich},
and the new physics
can make significant one-loop contribution to the $Zb\bar b$ vertex.
These contributions are of two types: self-energy corrections to the
external $b$-quark, and vertex corrections in the form of triangle diagrams.
However, they are significant
only if they involve a $t$-quark in the loop.

A $2.2\sigma$ deviation does not prove the existence of new physics, and
there are some doubts about the experimental number; e.g., the experiments
may have taken into account $b\bar b$ pairs generated from a gluon radiated
off a light quark, in which case actual $R_b$ will be smaller. Another
value quoted for $R_b$ is $0.2192\pm 0.0018$ \cite{glasgow},
in which case the deviation
is only $2\sigma$. In any case, the result encourages those models
which predict a positive deviation, e.g., a large parameter space in
the Minimal Supersymmetric Standard Model (MSSM),
or models with extra gauge bosons etc. On the other hand, it puts
very tight constraint on those models which predict the deviation to be
negative --- examples are the two-Higgs doublet model \cite{boulware},
and, as  will be
shown, the triplet model of Georgi and Machacek \cite{georgi2}.
But first we will show
our results for a general Higgs sector.

A very  concise formulation will essentially follow the notation set by
Boulware and Finnell \cite{boulware}. Let the
deviation of $R_b$ from its SM value be denoted by $\delta R_b$. Thus, we
have
\be
\delta R_b=0.1718\Delta
\ee
where the factor $\Delta$ contains the non-oblique one-loop effects
--- the oblique parts are already known to have negligible contribution.
The charged Higgs coupling to left-handed $b$-quarks is proportional
to $m_t$ while that to right-handed $b$-quarks is proportional to $m_b$.
Thus, the production of left-handed $b$-quarks is strongly favoured.
Since the same is true for the tree-level case, it will not cause any
significant change to the electroweak asymmetries.

In the limit $m_b\r 0$, the effects of new physics can
be introduced through a change in the vertex factors for the $Zb\bar b$
coupling:
\bea
{v^b_L}'&=&v^b_L+{g^2\over 16\pi^2}F_L(p^2,m_t),\\
{v^b_R}'&=&v^b_R+{g^2\over 16\pi^2}F_R(p^2,m_t),
\eea
where $p$ is the four-momentum of the $Z$ boson. Also, the right- and
left-handed couplings of $b$ (and $t$) quarks with $Z$ at the tree-level
are given by
\bea
v^b_R&=&{1\over 3}\sin^2\tw \\
v^b_L&=&-{1\over 2}+{1\over 3}\sin^2\tw \\
v^t_R&=&-{2\over 3}\sin^2\tw \\
v^t_L&=&{1\over 2}-{2\over 3}\sin^2\tw.
\eea
The form factors $F_L$ and $F_R$ can be written as
\be
F_{L,R}=\sum_{i=1}^4F^i_{L,R}
\ee
where the individual form factors $F^1$, $F^2$, $F^3$ and $F^4$ receive
contributions from the set of diagrams in Fig. 1(a), (b), (c) and (d)
respectively.

Both for models 1 and 2 discussed in Section 2, the $i$-th charged
Higgs couples with $b_L$ with the strength
\be
\lambda^i_L={g\over\sqrt{2}m_W}m_t{\alpha_{i1}\over \alpha_{11}}
\ee
whereas for the right-handed $b$, the expression is model-dependent:
\bea
\lambda^i_R=-{g\over\sqrt{2}m_W}m_b{\alpha_{i1}\over \alpha_{11}}\ \
{\rm (model~1)}\\
\lambda^i_R=-{g\over\sqrt{2}m_W}m_b{\alpha_{i2}\over \alpha_{12}}\ \
{\rm (model~2)}.
\eea
For model 1, $\lambda^i_R\ll \lambda^i_L$, and the same is true for
model 2 unless $\alpha_{i2}/\alpha_{12}\gg\alpha_{i1}/\alpha_{11}$.
Thus, even at the one-loop level, the produced $b$'s are dominantly
left-handed and the electroweak asymmetries are little affected. Also,
it is a relatively safe assumption to neglect the change in $v^b_R$,
which is roughly proportional to $(\lambda^i_R)^2$; the error introduced
is of the order of $(m_b/m_t)^2$, or $0.1\%$.

So $F^i_R\approx 0$, and $F^i_L$s are given by
\bea
F^1_L&=&\sum_{i=2}^nB_1(m_t,m_{H_i^+})v_L^b[\lambda^i_L]^2,\\
F^2_L&=&\sum_{i=2}^n\Big[[-m_Z^2(C_{11}+C_{23})(m_t,m_{H^+_i},m_t)
-{1\over 2}+2C_{24}(m_t,m_{H^+_i},m_t)]v^t_R\nonumber\\
&{}&+m_t^2C_0(m_t,m_{H^+_i},m_t)v^t_L\Big][\lambda^i_L]^2,\\
F^3_L&=&\sum_{i,j,k=1}^n\Big[-2C_{24}(m_{H^+_i},m_t,m_{H^+_j})\Big]
\alpha_{ik}\alpha_{jk}(\cos^2\tw-{1\over 2}Y_k)\lambda^i_L\lambda^j_L
\nonumber\\
&{ }&+\sum_{k=1}^n2C_{24}(m_W,m_t,m_W)(\cos^2\tw-{1\over 2}Y_k)(\alpha_{1k}
)^2[\lambda^1_L]^2,\\
F^4_L&=&\sum_{i=2}^n\sum_{k=1}^n-{1\over 2}\cos\tw\lambda^i_L\alpha_{ik}
(f^c_kv^c_k+f^r_kv^r_k)\nonumber\\
&{ }&\times[C_0(m_t,m_W,m_{H^+_i})+C_0(m_t,m_{H_i^+},m_W)],
\eea
where $B$ and $C$'s are the well-known two- and three-point functions
first introduced by Passarino and Veltman \cite{passarino},
and $f^c_k$ and $f^r_k$ are defined in eq. (16). The loop amplitudes are
evaluated with dimensional regularisation and $\overline{MS}$ subtraction
scheme, and the sum, $F_L$, is free of all divergences.

The absolute sign of $F^1_L$ and $F^2_L$ are straightforward: the first
one is negative while the second one is positive. $F^3_L$ and $F^4_L$ require
a more careful treatment.

What we are planning to do is to place a lower bound on the mass of the charged
Higgs(es). For that, the most profitable scheme is to take all physical charged
Higgses to be degenerate (or nearly degenerate) in mass. Otherwise, the
$C_{24}$
function will be dominated by the lowest mass eigenstate; the bound may be
slightly weakened at the cost of moving all other charged scalars to the heavy
mass regime. As the mass spectra cannot be investigated without complete
specification of the potential, this is the best that one can achieve in a
general treatment. This allows us to take the $C_{24}$ function to be nearly
a constant, and using the unitarity of the $\alpha$ matrix, one can recast
$F^3_L$ as
\bea
F^3_L&=&{g^2\over m_W^2}\Big[-{1\over (\alpha_{11})^2}
C_{24}(m_{H^+},m_t,m_{H^+})(\cos^2\tw-{1\over 2})\nonumber\\
&{ }&+C_{24}(m_W,m_t,m_W)\sum_{k=1}^n(\alpha_{1k})^2
(\cos^2\tw-{Y_k\over 2}) \Big].
\eea
The $C_{24}$ function is negative, so the first term on the right-hand
side of eq. (58)
is positive. The second term is positive for $Y_k\leq 1$, and negative for
$Y_k>1$ (we assume only integer $Y$ for scalar multiplets). Thus, for
complex doublets and real non-doublets, $F^3_L$ is positive definite.

However, here one must note a point. The constraints from $\zbb$ are
meaningful only if $\rho$ is forced to unity at tree-level. Otherwise, the
non-doublet VEVs are required to be so much smaller
than the doublet VEVs (from the
experimental constraint on $\rho$) that the doublet-nondoublet mixing angle
$\theta_H$ is very small, and $\zbb$ does not put any significant
constraint on $m_{H^+}$.

The tree-level value of $\rho$ can be forced to unity in models with
doublet and/or singlet scalar multiplets, or in models where the effects
of `bad' representations cancel out. In these models, all $\alpha_{1k}$'s
are specified from the definition of the charged Goldstone. If $m_{H^+}$
is of the same order of $m_W$, both the $C_{24}$ functions on the right-hand
side of eq. (58) are approximately same. In this case, $F^3_L$ {\em always}
turns out to be positive and proportional to $\tan^2\theta_H$. Even for
a larger $m_{H^+}$, $F^3_L$ is always positive. Also, these models necessarily
imply $F^4_L=0$. These two important results are proved in Appendix 2. Thus,
$F_L=F^1_L+F^2_L+F^3_L$, which turns out to be positive for $m_{H^+}
\leq 1$ TeV. $\Delta$ in eq. (43) is given by
\be
\Delta={g^2\over 16\pi^2}{2v_L^b\over (v^b_L)^2+(v^b_R)^2}F_L,
\ee
and thus $\delta R_b$ is negative, thus tightly constraining the parameter
space for these models. The lower bound on $m_{H^+}$ is inversely
proportional to $\tan^2\theta_H$; if $\alpha_{11}=\cos\theta_H=1$, all
other $\alpha_{1k}$'s are zero, and $F_L$ is also zero; thus there is no bound
on $m_{H^+}$. It is also evident that the approximate magnitude of the
bound is independent of the number of `bad' representations.

A word about $F^4_L$, though it may be nonzero  only
in some contrived models (e.g., one with a $T=3,Y=4$ multiplet). Using the
same logic as before, we can take $C_0$ to be a constant, and $F^4_L$ can
approximately be written as
\bea
F^4_L&=&-{gm_t\over 2\sqrt{2}m_W}\cos\tw C_0(m_t,m_W,m_{H^+})\Big[ (\cos\tw
-\sec\tw)v\nonumber\\
&{ }&-{1\over v^2}\sum_{k=1}^n\{(v^c_k)^2f^c_k+(v^r_k)^2f^r_k\}\Big]
\eea
where $m_W^2=g^2v^2/2$. $C_0$ being positive, the first term in the RHS
is positive and gives negative $\delta R_b$; the second term adds in the
same direction if $Y_k\leq 1$.

The above discussion reflects a rather interesting complementarity.
Scenarios which do not ensure $\rho=1$ at tree-level cannot be restricted
unequivocally using $R_b$; however, such scenarios are severely constrained
by the value of $\rho$ itself. If, on the other hand, the tree-level value
of $\rho$ is somehow fixed at unity through some additional symmetry
in a complicated scalar sector,  then these very conditions which ensure
$\rho=1$ make it possible to always constrain the scalar sector through
$R_b$. Thus the near-unity of $\rho$, coupled with precision
measurement of $R_b$, limits the parameter space of an arbitrarily
extended Higgs structure.

\bigskip

As a nontrivial example of a model with custodial symmetry preserving
scalar sector, we again consider the triplet model. In this model, $F^4_L=0$
and the rest $F^i_L$'s are all proportional to $(\lambda^3_L)^2$. Thus, only
$H^+_3$ can be constrained --- $H^+_5$ does not couple with the fermions
and do not contribute to the loop-amplitude. $\lambda^3_L$ being proportional
to
$\tan\theta_H$, we take two representative values: $\sin\theta_H=0.5$ and
$\sin\theta_H=0.8$. $m_t$ is varied over the range $176\pm 13$ GeV. The
resulting $\delta R_b$'s are shown in Figures 2 and 3 respectively.

For $\sin\theta_H=0.5$, the lower bound on $m_{H_3^+}$ is too small to be
of any significance. Even direct
experiments put a better bound. For models with
$\rho_{tree}\not = 1$, $\sin\theta_H$ is even smaller. For $\sin\theta_H>0.8$,
the lower bound on $m_{H^+_3}$ is almost 1 TeV.
Thus we can say that $\sin\theta_H=0.8$ is the
maximum mixing allowed in this model, unless one wants to have
a scalar with mass above 1 TeV. Note that this is about $3-4$ times
stronger than the bound derived by Gunion {\em et al} \cite{gunion}
from FCNC processes, and also relatively free from hadronic uncertainties.

\bigskip\bigskip

\centerline{\bf 5. Higgs Production}
\bigskip

The production of non-standard Higgs bosons in both hadronic and $e^+e^-$
colliders has been extensively discussed in the literature.
In particular, the production of scalars with exotic charges (like $H^{++}$)
has received considerable attention \cite{vega,lusignoli}.
In this section, we include a brief
discussion on the production of non-standard scalars belonging to
arbitrary multiplets in $e^+e^-$ collisions.

\bigskip
\centerline{\sf A. Neutral Scalar Production}
\bigskip

The production of a neutral scalar that can couple to a pair of gauge bosons
can always take place through the Bjorken process $e^+e^-\r Z
\r Z^{\ast}H^0_{Si}$ in a $Z$-factory (or $e^+e^-\r Z^{\ast}
\r ZH^0_{Si}$ in a higher energy $e^+e^-$ machine) \cite{hunter}.
In the general
case, the scalars produced in this manner will decay into a pair of fermions
(say, $b\bar b$), and the final state will consist of four fermions, two of
which will have an invariant mass equal to the respective scalar mass.
However, many extended models (aimed at keeping $\rho_{tree}=1$) possess
some additional symmetries that forbid the interaction of some of these scalars
with fermions. In such cases, those scalars will decay either into four
fermions, induced by a pair of real or virtual gauge bosons, or into two
fermions via loops. Assuming that the former mode dominates, the ratio of
the contributions to $Z\r 4f$ and $Z\r 6f$ channels from all
the scalar mass eigenstates is given by
\be
{\Gamma(Z\r\sum_{i=1}^mZ^{\ast}H^0_{Si}\r 4f)\over
\Gamma(Z\r\sum_{i=m+1}^nZ^{\ast}H^0_{Si}\r 6f)}
={\sum_{i=1}^m\sum_{k,l}G_i\beta_{ik}\beta_{il}v_k^2v_l^2Y_kY_l\over
\sum_{i=m+1}^m\sum_{k,l}G_i\beta_{ik}\beta_{il}v_k^2v_l^2Y_kY_l}, \ee
where it has been assumed only that the physical scalar states, from
$i=1$ to $m$, couple to fermions. $G_i$ is given by
\be
G_i=\int_{x_0}^{x_1} ~g(x)~dx,
\ee
with
\be
g(x)={(1-x+{x^2\over 12}+{2\over 3}y^2)(x^2-4y^2)^{1/2}\over (x-y^2)^2}
\ee
where
\be
y=m_{H^0_{Si}}/m_Z.
\ee

It is clear from above that the maximum value (with all degenerate
scalars) of the four-fermion signals via Bjorken process is that for the
single Higgs doublet in the SM. Also, the neutral components of real
scalar multiplets do not contribute to any of the two kinds of signals.
The above result can be extended in a straightforward way to
$e^+e^-\r Z^{\ast}\r \sum_i ZH^0_{Si}$ at a higher
energy.

\bigskip
\centerline{\sf B. Neutral Pseudoscalar Production}
\bigskip

Pseudoscalars are produced through the mechanism $Z(Z^{\ast})\r
H^0_{Si}H^0_{Pj}$. The process and the ensuing signals are closely
analogous to the signal of the pseudoscalar in two-Higgs doublet models
\cite{pocsik}.
The only interesting difference might occur if some pseudoscalars do
not have tree-level couplings to fermions. In such cases, the
pseudoscalar decays through channels involving real or virtual gauge
bosons or scalars.

\bigskip
\newpage
\centerline{\sf C. Charged Scalar Production}
\bigskip

While the mode $Z(Z^{\ast})\r H_i^+H_j^-$ is still open,
a new avenue for charged scalar production opens up when higher
Higgs representations are present. The $H^{\pm}_iW^{\mp}Z$ vertex
in a general scenario leads to the process $Z(Z^{\ast})\r
H_i^{\pm}{W^{\ast}}^{\mp}(W^{\mp})$. The width for $Z\r
H_i^{\pm}{W^{\ast}}^{\mp}\r H_i^{\pm}(p_3)f_1(p_1)
\bar{f_2}(p_2)$ is given by
\be
\Gamma_i={\cal G}_i\sum_{k,l}t_kt_l\alpha_{ik}\alpha_{il}
\ee
where
\be
t_k=(\cos\tw-Y_k\sec\tw)\sqrt{n_k-1}v_k
\ee
and
\be
{\cal G}_i={1\over 8\pi^2m_Z} \int~dE_1dE_2|{\cal M}|^2,
\ee
$|{\cal M}|^2$ being the relevant squared matrix element. the signal of
a charged Higgs produced in this way will consist chiefly of two- or
four-fermion decay modes, depending upon whether the $H^{\pm}_i$
couples to fermions or not at the tree-level. Again, the latter
possibility is often the consequence of symmetries imposed on the scalar
sector to maintain $\rho_{tree}=1$. Thus a real or virtual $Z$ in
$e^+e^-$ machines will give rise to four- or six-fermion signals as a
consequence of charged Higgs production through the $H_i^{\pm}W^{\mp}Z$
vertex. Again, assuming that $(n-m)$ out of $n$ singly charged scalars
do not have tree-level fermionic coupling, one obtains
\be
{\Gamma(Z\r 4f)\over \Gamma (Z\r 6f)}
={\sum_{i=2}^m\sum_{k,l}{\cal G}_i\alpha_{ik}\alpha_{il}t_kt_l\over
\sum_{i=m+1}^n\sum_{k,l}{\cal G}_i\alpha_{ik}\alpha_{il}t_kt_l},
\ee
where $i=1$ has been left out of the sum in order to separate out the
charged Goldstone field. In the formula above, any $H^{\pm}_i$ will
have $H_i^{\pm}f_1\bar{f_2}$ interaction if\\
\noindent (i) $\alpha_{i1}\not =0$ in the case where only the doublet
$\Phi_1$ gives masses to all the fermions, and\\
\noindent (ii) $\alpha_{i1},\alpha_{i2}\not = 0$ in the case when $\Phi_1$
and $\Phi_2$ are responsible for the masses of up- and down-type fermions
respectively.

The observable consequences of the $H_i^{\pm}W^{\mp}Z$ vertex in both
LEP-1 and higher energy machines have been discussed in detail in the
context of the triplet model \cite{hontka,roger}.
It has also been shown \cite{hontka} that in cases
where the $H_i^{\pm}$ does not couple to fermions, its dominant
decay mode is the tree-level one into four fermions over most of the
parameter space.

We illustrate in Fig. 4 the branching ratio for $Z\r H_i^{\pm}
\ell\bar{\nu_l}$ as a function of the $H_i^{\pm}$ mass in the triplet model.
It is clear from the graphs that a considerable range of the parameter
space of $m_{H^+}$ and $\theta_H$, the doublet-triplet mixing angle,
can be constrained from the existing experiments.

\bigskip\bigskip

\centerline {\bf 6. Conclusions}
\bigskip

We have discussed the phenomenology of a general scenario with an
arbitrary combination of real and complex Higgs multiplets in arbitrary
representations of $SU(2)$. We have seen that with some very modest
assumptions, most of the interactions and formulae in such a scenario
can be obtained in rather simplified and physically transparent form.
When the multiplets are such that they do not ensure $\rho=1$ at the
tree-level, the experimental value of $\rho$ itself is the most
stringent constraint on them. On the other hand, for those models
where $\rho_{tree}=1$ is ensured by suitable contrivance, the precision
measurement of $\Gamma (\zbb)$ strongly constrains the parameter space.
The consideration of unitarity sum rules in longitudinal gauge boson
scattering can also yield interesting relationships among the
parameters in a general structure.

On the whole, an extended scalar structure in the electroweak symmetry
breaking scheme has  a rich phenomenology and deserves unbiased
scrutiny. The search for such `exotic' Higgs particles should therefore
be given due priority in all the present and future experiments.

\bigskip\bigskip
\centerline{\sf Acknowledgements}
\bigskip

A.K. and B.M. acknowledge the warm hospitality of the Mehta Research
Institute, Allahabad, and the Saha Institute of Nuclear Physics, Calcutta,
respectively. They also thank  B. Dutta-Roy and R.T. Adhikari
for useful discussions
and A.K. Dutt- Mazumder for help with some of the figures. B.M. is
grateful to M. Nowakowski for illuminating discussions on exotic Higgs
models.

\newpage

\setcounter{equation}{0}
\appendix
\renewcommand{\theequation}{1.{\arabic{equation}}}

\centerline{\bf Appendix 1}

In this appendix, we present the vertex factors for a general
assortment of scalar multiplets. We display only those vertex
factors which do not depend explicitly on the form of the scalar
potential. Thus, the three-scalar and four-scalar vertices are
omitted.

\bigskip

{\bf 1.} Two gauge bosons and one scalar:
\bea
ZZH_{Si}^0\ &:&\ {ig^2\over\sqrt{2}\cos\tw}g_{\mu\nu}\sum_k
\beta_{ik}v_kY_k^2,\\
W^+W^-H_{Si}^0\ &:&\ {ig^2\over\sqrt{2}}g_{\mu\nu}\sum_k
\beta_{ik}(Y_kv_k^c+{1\over 4}(n_k^2-1)v_k^r),\\
W^+ZH_{i}^-\ &:&\ {ig^2\over\sqrt{2}\cos\tw}g_{\mu\nu}\sum_k
\alpha_{ik}(\fcfr ),\\
W^+W^+H_{i}^{--}\ &:&\ {ig^2\over{2}}g_{\mu\nu}\sum_k
\sigma_{ik}(\gcgr ),\eea
where $f_k^c$, $f_k^r$, $g_k^c$ and $g_k^r$ are defined in eqs.
(16), (18) and (19).

\bigskip

{\bf 2.} Two scalars and one gauge boson:\\
($p_1$ and $p_2$ are respectively incoming and outgoing momenta at
the vertex)
\bea
ZH_i^+H_j^-\ &:&\ -{ig\over 2\cos\tw}(p_1+p_2)_{\mu}
[2\cos^2\tw\delta_{ij}-\sum_k\alpha_{ik}\alpha_{jk}Y_k], \\
ZH_{Si}^0H_{Pj}^0\ &:&\ {g\over 2\cos\tw}(p_1+p_2)_{\mu}
\sum_k\beta_{ik}\gamma_{jk}Y_k, \\
ZH_i^{++}H_j^{--}\ &:&\ -{ig\over 2\cos\tw}(p_1+p_2)_{\mu}
[4\cos^2\tw\delta_{ij}-\sum_k\sigma_{ik}\sigma_{jk}Y_k], \\
W^-H_i^{+}H_{Sj}^0\ &:&\ -{ig\over \sqrt{2}}(p_1+p_2)_{\mu}
\sum_k\alpha_{ik}\beta _{jk}(h_k^c+h_k^r), \\
W^-H_i^{+}H_{Pj}^0\ &:&\ {g\over \sqrt{2}}(p_1+p_2)_{\mu}
\sum_k\alpha_{ik}\gamma_{jk}(h_k^c+h_k^r), \\
W^-H_i^{++}H_j^{-}\ &:&\ -{ig\over \sqrt{2}}(p_1+p_2)_{\mu}
\sum_k\sigma_{ik}\alpha_{jk}(q_k^c+q_k^r), \eea
where
\bea
h_k^c=\sqrt{n_k-1}\ &;&\ h_k^r={1\over 2}\sqrt{n_k^2-1};\\
q_k^c=\sqrt{2(n_k-2)}\ &;&\ q_k^r={1\over 2}\sqrt{n_k^2-9}.\eea

In general, in the weak basis, if $W$ couples with two members of charges
$j$ and $j+1$ in a particular multiplet of dimension $n$, then the vertex
factor is given by
\bea
W^-\phi^{+(j+1)}\phi^{-j}\ &:&\ -{ig\over\sqrt{2}}(p_1+p_2)_{\mu}
\sqrt{(n-j-1)(j+1)} ~~{\rm (complex)},\nonumber\\
&{ }&\\
W^-\phi^{+(j+1)}\phi^{-j}\ &:&\ -{ig\over 2\sqrt{2}}(p_1+p_2)_{\mu}
\sqrt{n^2-(2j+1)^2} ~~{\rm (real)},\eea
which can be easily translated to the mass basis if the mixing matrices
are known.

\bigskip

{\bf 3.} Two gauge bosons and two scalars:\\
These are presented below in the weak basis.
\bea
W^+W^-\phi^{+Q}\phi^{-Q}\ &:&\
-ig^2g_{\mu\nu}[T(T+1)-(Q-{Y\over 2})^2], \\
W^+W^+\phi^{-(Q+2)}\phi^{Q}\ &:&\
-i{g^2\over 2}g_{\mu\nu}(T^+)^2, \\
W^+W^+\phi^{--}\phi^{0}\ &:&\
-i{g^2\over 2}g_{\mu\nu}g_k^c ~~(g_k^r~{\rm for~ real}), \\
AA\phi^Q\phi^{-Q}\ &:&\
-{ig_{\mu\nu}}e^2Q^2, \\
ZZ\phi^Q\phi^{-Q}\ &:&\
-ig_{\mu\nu}{g^2\over 4\cos^2\tw}(2Q\cos^2\tw-Y)^2,\\
AZ\phi^Q\phi^{-Q}\ &:&\
-ig_{\mu\nu}{eQg\over \cos\tw}(2Q\cos^2\tw-Y),\\
AW^-\phi^{Q+1}\phi^{-Q}\ &:&\
-ig_{\mu\nu}{eg(2Q+1)\over \sqrt{2}}T^+,\\
ZW^-\phi^{Q+1}\phi^{-Q}\ &:&\
-ig_{\mu\nu}{g^2\over\sqrt{2}}~
\big[ {\cos\tw (2Q+1)}+Y{\cos 2\tw\over\cos \tw}\big]~T^+,\nonumber\\
&{ }&\eea
where $T^+$ is a shorthand notation for $<\phi^{Q+1}|T^+|\phi^Q>$.

\newpage

\setcounter{equation}{0}
\appendix
\renewcommand{\theequation}{2.{\arabic{equation}}}

\centerline{\bf Appendix 2}

Here we will state and prove two theorems about the vertex
correction to the process $\zbb$, in which singly charged scalars of
arbitrary representation of $SU(2)$ take part.

\bigskip

{\em Definition:} By $\rho_{tree}=1$ models, we mean those models whose
scalar sector consists of (i) either complex doublets (and singlets), which
guarantee $\rho=1$ at tree-level, or (ii) a set of `bad' representations
whose effects on $\rho-1$ at tree-level cancels out, and which have been
constrained according to our prescription laid down in Section 2, apart
from the usual doublet. There may be more than one doublet, and singlets too.
Moreover, we confine ourselves to those complex multiplets for which
$T^-v_k=0$.

\bigskip

{\em Theorem 1:} $F^3_L$, as defined in eq. (58), is always positive for
$\rho_{tree}=1$ models.

{\em Proof:} The proof will be given for case (ii) of the above definition
only, as the proof for case (i) follows trivially.

Let us take the VEV of the doublet $\Phi_1$, which gives mass to the top
quark, to be $v_d$. We also consider one complex $m$-plet, $\Phi_2$,
with VEV=$v_m$ ($T=Y/2$), and one real $m$-plet, $\Phi_3$, with VEV=
$v'_m=\sqrt{4T-2/T+1}v_m$.

The Goldstone boson is defined as
\be
G^+={g\over\sqrt{2}m_W}\sum_i[(T^+v_i)^{\dag}\phi_i^+]=\sum_i
\alpha_{1i}\phi_i^+,
\ee
as we take $T^-v=0$, without any loss of generality. Also,
\bea
T^+v&=&\sqrt{2T}v\ \  \ {\rm (complex)},\nonumber\\
&=&\sqrt{T(T+1)}v \ \ \ {\rm (real)}. \eea
Thus,
\bea
(\alpha_{11})^2&=&{g^2\over 2m_W^2}v_d^2,\\
(\alpha_{12})^2&=&{g^2\over 2m_W^2}2Tv_m^2,\\
(\alpha_{13})^2&=&{g^2\over 2m_W^2}T(4T-2)v_m^2.\eea
Therefore,
\bea
\sum_k(\alpha_{1k})^2(\cos^2\tw-{1\over 2}Y_k)&=&
{g^2\over 2m_W^2}(v_d^2+4T^2v_m^2)(\cos^2\tw-{1\over 2})\nonumber\\
&=&\cos^2\tw-{1\over 2}, \eea
as $g^2(v_d^2+4T^2v_m^2)=2m_W^2$.

$C_{24}$ being always negative for the arguments used in the definition of
$F_L^3$, and $\cos^2\tw > 1/2$ and $\alpha_{11}^2\leq 1$, it follows
that $F_L^3$ is positive definite. The same proof follows for $F_R^3$, and
also for models with more than one set of `bad' representations and/or
more than one doublets.

\bigskip
\newpage

{\em Theorem 2:} $F^4_L=0$ for $\rho_{tree}=1$ models.

{\em Proof:} We take the same assortment of scalars. $F^4_L$ is proportional
to
\be
\sum_{i=2}^n\sum_{k=1}^n{\alpha_{i1}\over\alpha_{11}}\alpha_{ik}
(\fcfr ), \ee
which can be rewritten as
\be
{1\over\alpha_{11}}f_k^1v_d-\sum_{k=1}^n\alpha_{1k}(\fcfr ).
\ee
Putting the values of $f^c_k$, $f^r_k$, $v^c_k$ and $v^r_k$, as shown in
Theorem 1, we find that the expression {\em vanishes identically}.

In the proof of the above two theorems, we have assumed that $m_{H^+}$
is of the same order of $m_W$. The theorems are valid for $m_{H^+}
\sim 1$ TeV, beyond which the perturbative unitarity breaks down.

\newpage

\centerline{\large\bf Figure Captions}

\begin{enumerate}

\item Feynman diagrams involving charged scalars
contributing to the one-loop correction to
the $\zbb$ vertex.

\item $\delta R_b$ plotted against the charged Higgs mass for
$\sin\theta_H=0.5$. The solid, dotted and dashed lines correspond to
$m_t=163,176,189$ GeV respectively.

\item Same as in Figure 2, but with
$\sin\theta_H=0.8$.

\item Branching ratio for $Z\r H^+\ell\bar{\nu_l}$ plotted against
$m_{H^+}$. The solid and dashed lines correspond to $\sin\theta_H
=0.1$ and 0.8 respectively.

\end{enumerate}
\end{document}